# Coherent Unipolar Cherenkov and Diffraction Radiation Generated by Relativistic Electrons

G. Naumenko*, M. Shevelev, K.E. Popov

Tomsk Polytechnic University, 634004, Lenin ave. 30, Tomsk, Russia

For usual radiation the integral of the electric field strength over the time is zero. First time the possibility of unipolar radiation generation has been considered theoretically in the paper [EG.Bessonov. Preprint No 76 of the Lebedev Physical Institute of the Academy of Science of the USSR, Moscow, 1990]. According to this work the unipolar radiation is radiation for which the integral of the electric field strength over the time differs significantly from zero. Futher a number of theoretical articles have been devoted to this problem mainly in respect to a synchrotron radiation. However up to now there are no experimental investigations of this phenomenon. In this paper we present the results of experimental observation of the unipolar Cherenkov and diffraction radiation generated by relativistic electrons in millimeter wavelength region. For this purpose the detector sensitive to the selected direction of the electric field strength has been elaborated. We observed the coherent Cherenkov radiation and backward diffraction radiation of bunched electron beam when the electrons moving near the targets. The partial unipolarity effect has been registered for the Cherenkov radiation and the almost full unipolarity has been observed for the backward diffraction radiation.

## I. INTRODUCTION

The first mention of unipolar radiation belongs to Bessonov [1], who theoretically investigated unipolar pulses named as "strange electromagnetic waves", which satisfy the condition $\int_{-\infty}^{\infty} E(t)dt \neq 0$. The Fourier component of electric field can be written as $E_\omega = \int e^{i\omega t} E(t)\,dt$. One can show that if $\int E(t)dt = 0$, then in classical (non-quantum) approximation $E_{\omega \to 0} = 0$, and vice versa. Thus, relationship $\dfrac{E_{\omega \to 0}}{\int |E(t)|dt}$ in indicated approximation can be regarded as a normalized criterion for radiation unipolarity.

This work caused a stormy theoretical study of this problem. In [2] the possibility of unipolar pulse radiation in the context of Cherenkov radiation when medium is excited by ultrashort pulses at superluminal velocity was considered. The possibility of an unipolar pulse generation in nonlinear media in terahertz and optical electromagnetic ranges was considered theoretically in number of articles [2,3,4]. Also, the existence of unipolar (half-cycle) solutions of Maxwell equations has been considered in [5-9]. The possibility of generating an unipolar radiation of relativistic electrons in a magnetic field has been theoretically shown in [1,10,11]. However, experimental studies devoted to the emission and detection of unipolar radiation did not carried out.

---

* Corresponding author.

In this paper we present the results of the first experimental investigations of the unipolar coherent Cherenkov radiation (ChR) and coherent backward diffraction (BDR) radiation generated by relativistic electrons, which travel in the vicinity of targets. According to [12], Cherenkov radiation can be generated when an electron beam passes near a dielectric target. This geometry is very similar to the geometry of diffraction radiation.

Diffraction radiation is the radiation of electrons moving near an edge of target without crossing it. In [13,14] one may find the theoretical consideration of the backward diffraction radiation of relativistic electrons from a conducting semiplane. These conditions are very close to our experimental one. In [13] the authors obtained the solution for BDR vector potential in Fourier presentation (a schematic view of radiation geometry shown in figure 1):

$$A_x(R) = \frac{2\pi e^{i\omega R}}{R} j_x(k_0, q_0),$$

where $R$ is the distance from the target to the observation point. In system, where $x = R\sin\psi\cos\varphi$, $y = R\sin\psi\sin\varphi$, $z = R\cos\psi$, $k_0 = -\omega\sin\psi\cos\varphi$, $q_0 = -\omega\cos\psi$.

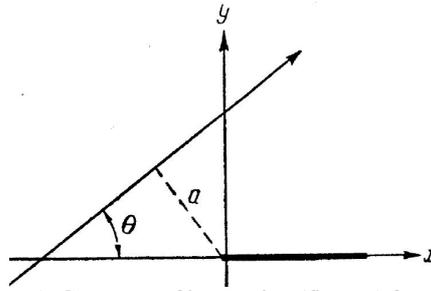

**Figure 1.** Geometry of interaction (figure 1 from [13]).

Expression for current $j_x(k,q)$ has complicated view and one may find it in [13]. The vector potential $A_x$ corresponds to the horizontal component of polarization. After the back Fourier transform we obtain the expression for the time dependence of the horizontal polarization component of the electric field strength of BDR

$$E_x(t) = \int e^{i\omega t} \omega \cdot A_x(R) d\omega.$$

This dependence for experimental conditions (see paragraph II) is shown in figure 2.

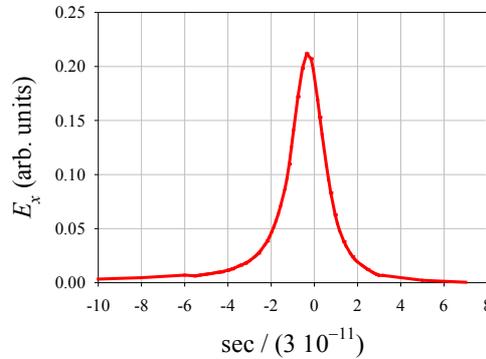

Figure 2. The time dependence of the horizontal polarization component of the BDR electric field strength.

As is seen from figure 2 the BDR is strongly unipolar.

In the pseudo-photon approximation [15,16] the ChR of relativistic electrons in considered geometry is the refracted electron field on the target surface and the BDR of relativistic electrons from a conducting target is the reflection of the electron field from the target. For the BDR geometry (figure 4) the horizontal polarization component of the reflected electric field of electrons keeps the same direction and we can expect that BDR is unipolar.

## II. EXPERIMENTAL SETUP

### A. Layout of experiment

The experiment was carried out using the relativistic electron beam extracted from the microtron in Tomsk Polytechnic University. The beam parameters are listed in Table 1. Under these conditions, radiation from the electron bunch at wavelengths $\lambda > 9$ mm ($\nu \leq 33.3$ GHz) is coherent [17]. Since the process of radiation generation is coherent, the intensity of radiation increases by a factor of $N_e$, where $N_e$ is electron number in a bunch, and it can be measured by existing detectors at room temperature.

Table 1. Beam parameters

| | |
|---|---|
| Electron energy | 6.1 MeV ($\gamma = 12$) |
| Macro-pulse (train) duration | 4 $\mu s$ |
| Bunch length | $\sigma_z = 3 \pm 1$ mm |
| Bunch population | $10^8$ |
| Bunches in train | $10^4$ |
| Distance between bunches | $\Lambda = 114$ mm |
| Extracted beam size | $4 \times 2$ mm |

The measurement time is determined by the pulse duration $\tau = 4\,\mu s$ of the electron beam, which corresponds to the wavelength $\lambda = 2\pi c\tau = 3 \cdot 10^3$ mm, however, since we measure only the coherent radiation from the electron bunch, the minimum wavelength of the measured radiation is determined by the longitudinal bunch form factor and is equal $\lambda_{min} : 9$ mm.

The layouts of the experiments on ChR and BDR unipolarity are shown in figures 3 and 4 respectively. The detector is placed in the focus of the parabolic mirror with focus length $f = 151$ mm to provide the measurement of angular distribution of radiation in the far field zone (see [18]). The impact-parameters (the distances between the electron beam and the targets) in both schemes have been equal to 15 mm. The transversal beam size near the targets is $\sigma_{BS} = 15$ mm.

For generation of ChR we used the Teflon prism of dimensions $175 \times 175 \times 70$ mm. The $\theta$-scans for ChR (figure 3) has been performed by rotation of the flat mirror around the vertical axis. The measurements were carried out at the observation angle near the cherenkov angle for used target $\psi \approx 45^0$.

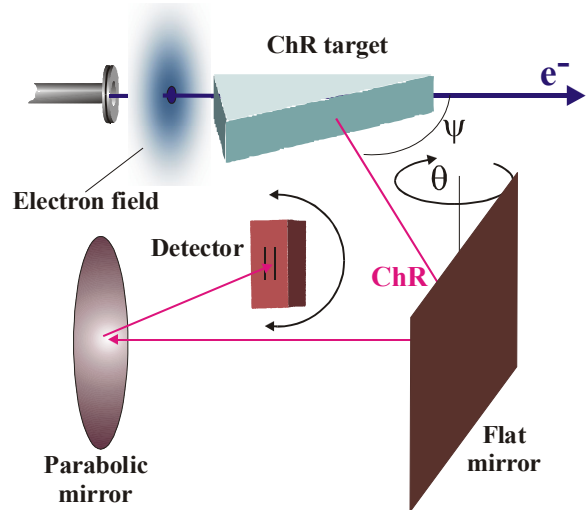

**Figure 3.** Layout of experiment on ChR unipolarity.

For BDR measurements (figure 4) the observation angle has been chosen $\psi = 90^0$. The target could be rotated around its vertical edge for a radiation orientation dependence measurement (so-called $\theta$-scan, where $\theta$ is the angle between target surface and electron beam direction). The distance from the target to the parabola is 300 mm and $11\,mm < \lambda < 14\,mm$. In this case, the formation length $l_f : \lambda/2\pi \approx 2\,mm$, which is much less than the distance from the target to the parabola.

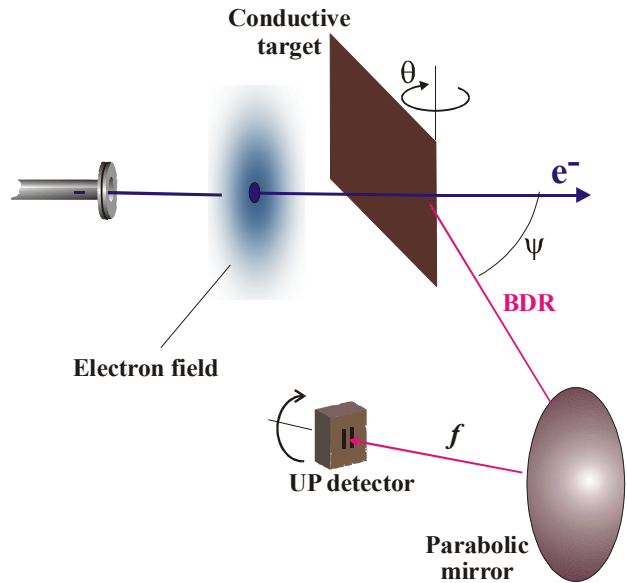

**Figure 4.** Layout of experiment on BDR unipolarity..

### B. Detector

To measure unipolar radiation, a detector based on the well-known technique applied for the surface current measurement in strip-line beam position monitors [19] has been developed. The scheme of detector is shown in figure 5. Surface currents induced by radiation in different directions are dispensed by microwave diodes to different channels: channel No 1 and channel No 2.

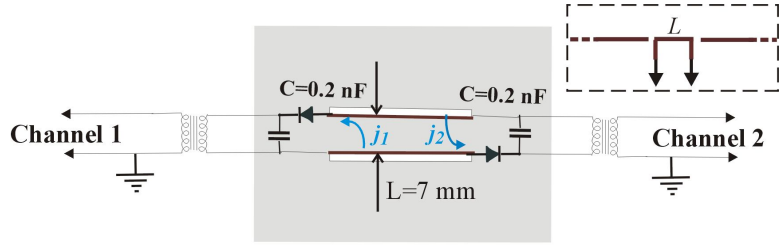

**Figure 5.** Scheme of detector. In insertion the profile of strip is shown.

The spectral efficiency of the developed detector according to [19] is shown in Fig. 7, where $v_0 = \frac{c}{4L} = 10.7\ GHz$ and $c$ is the speed of light. Spectral range of detection equals $\Delta v_{FWHM} = 10\ GHz$.
In the current experiment the measurements of the radiation orientation dependence were carried out for two positions of the detector, which are demonstrated in figure 8. Thus, different directions of the electric radiation field are detected by different channels.

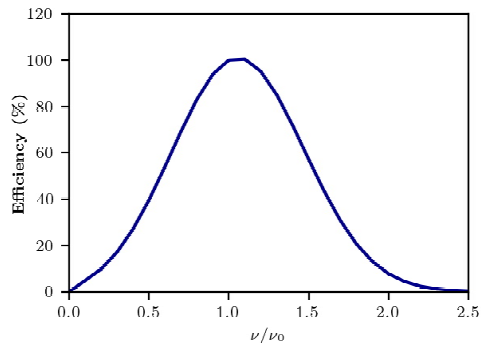

**Figure 7.** Spectral efficiency of detector

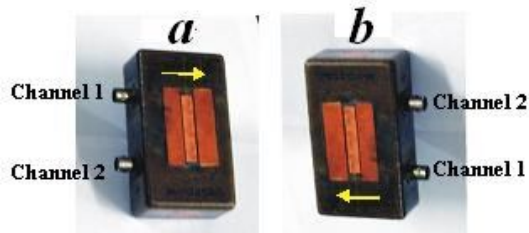

**Figure 8.** Indication of the detector position.

### III. MEASUREMENT RESULTS

Using technique described we measured the $\theta$-scans of ChR and BDR from channels No 1 and No 2 for the *a* and *b* detector positions as shown in figure 8. The results of the $\theta$-scans obtained after background subtraction are shown in figures 9 and 10 for BDR and in figures 11 and 12 for ChR. The background was measured without targets for each position of the detector. The statistical errors in presented arb. units are 0.0022 for channel No 1 and 0.0015 for channel no 2.

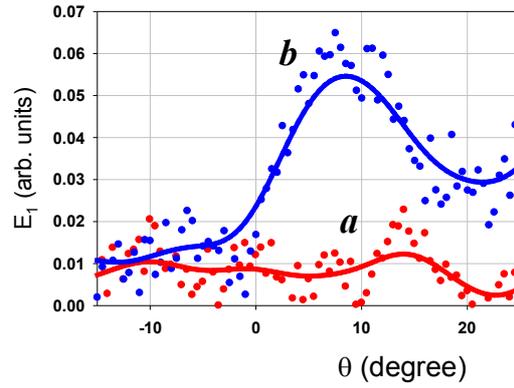

**Figure 9.** The $\theta$-scan of BDR from channel 1 for the *a* and *b* detector positions (see figure 8). The solid lines are the smoothed experimental data.

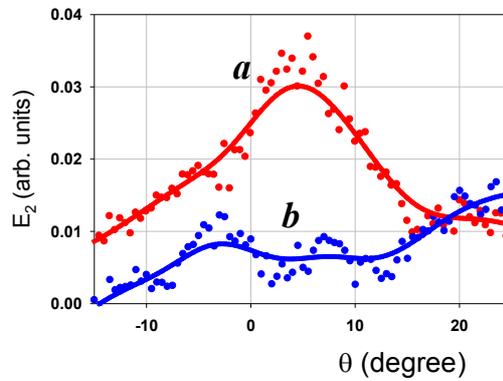

**Figure 10.** The $\theta$-scan of BDR from channel 2 for the *a* and *b* detector positions (see figure 8). The solid lines are the smoothed experimental data.

One can see in figures 9, 10 that the channels No 1 and No 2 demonstrate opposite results for the *a* and *b* detector positions. Obtained results suggest that the induced currents and, hence, the strength of the electric field of radiation have an allotted direction. We can argue, that BDR is almost completely unipolar one.

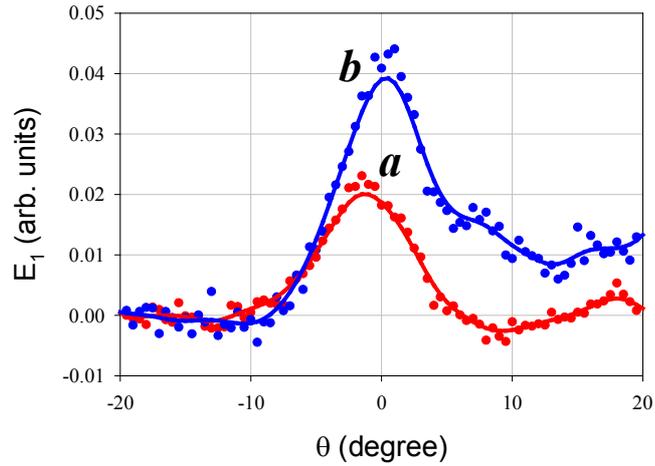

**Figure 11.** The $\theta$-scan of ChR from channel 1 for the ***a*** and ***b*** detector positions (see figure 8). The solid lines are the smoothed experimental data.

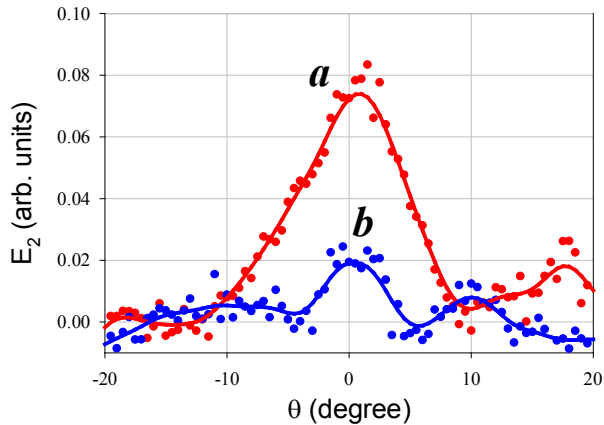

**Figure 12.** The $\theta$-scan of ChR from channel 2 for the ***a*** and ***b*** detector positions (see figure 8). The solid lines are the smoothed experimental data.

In case of ChR (figure 11,12) we see, that radiation is partly unipolar and $\dfrac{E_{\omega \to 0}}{\int |E(t)| dt}$ probably becmes significant.

### IV. CONCLUSION

In this paper we demonstrated that BDR is almost completely unipolar radiation and ChR is partly unipolar. The detector developed provides the possibility of detection of the unipolarity effect of radiation. Our next steps will be to assume:

a) Optimization of the parameters of the unipolar radiation detector. We should note that it is necessary to modify the detector including full matching of its elements such as antenna characteristics, microwave diodes and signal registration circuitry, as well as providing pre-amplification of the signal, in future experiments;

b) In [1,10,11] authors theoretical studied synchrotron radiation and demonstrated that it is unipolar. Thus, experimental investigation of unipolarity of synchrotron radiation is waited for;

c) From fundamental point of view the particular interest is the analysis of the bremsstrahlung unipolarity. According to [20], to calculate the radiation field in the low-frequency approximation, one can use the expression for the Liénard-Wiechert potentials:

$$\vec{E} = \frac{e}{c^2} \frac{1}{\left(R - \frac{\vec{v}\vec{R}}{c}\right)^3} \vec{R} \times \left(\left(\vec{R} - \frac{\vec{v}\vec{R}}{c}\right) \times \vec{a}\right),$$

where $e$ is the electron charge, $\vec{R}$ is the radius of the observation vector, $\vec{v}$ is the electron velocity, and $\vec{a}$ is the electron acceleration in the field of the nucleus, which for $\gamma \gg 1$, where $\gamma$ is the Lorentz factor, takes the form: $\vec{E} \approx -\frac{e\gamma^4}{Rc^2}\vec{a}$. Obviously, $\vec{a}$ is determined by which side and at what distance from the nucleus the electron moves. The last parameters can be fixed by using the bremsstrahlung photon tagging system (see figure 12).

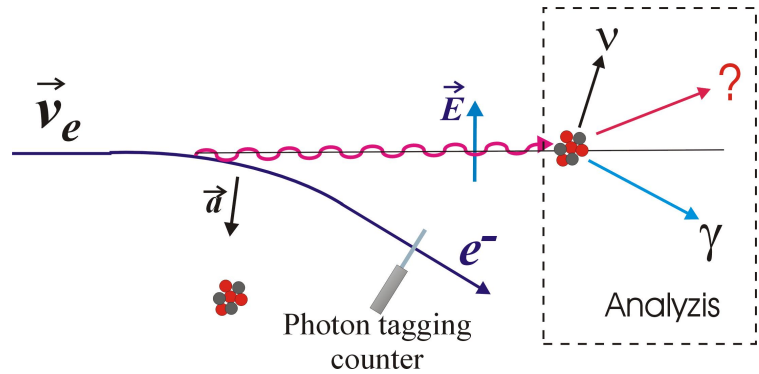

**Figure 12.** Scheme of the system of marking of bremsstrahlung photons and their analysis.

In this approximation, the direction and modulus of the vector $\vec{a}$ will be fixed and the radiation field will be unipolar. In this experiment the problem of determination of unipolarity bremsstrahlung radiation will arise. The problem is that in the quantum approximation the notion of photon unipolarity is absent, but as was shown above, in classical approximation a bremsstrahlung can be unipolar. If this fact takes place, then using the unipolar bremsstrahlung in a reaction which can be sensitive to the unipolarity, one can obtain an unexpected and unexplained asymmetry of reaction products in frame of quantum approximation. For the suggested experiment it is necessary for analysis to find a reaction which will be sensitive to unipolar radiation. In principle, we can consider the Tomson or the Compton scattering, if we can find the classical or quasi-classical approximation of this reaction. Observation of this effect has fundamental importance.

## ACKNOWLEDGMENTS

This work was supported by the Ministry of Education and Science of the Russian Federation (program "Science," basic part, No 3.8427.2017/8.9) and by the Competitiveness enhancement program of Tomsk Polytechnic University.

———————————————————